\documentclass[aps,prl,twocolumn,groupedaddress,showpacs,floatfix]{revtex4}
\usepackage{graphicx}
\begin{document}

\title{Colloid-polymer mixtures in the protein limit} \author{Peter
G. Bolhuis$^*$, Evert Jan Meijer$^*$ and Ard A. Louis$^\ddagger$}
\affiliation{$^*$Dept. of Chemical Engineering, University of
Amsterdam, Nieuwe Achtergracht 166, 1018 WV, Amsterdam, Netherlands\\
$^\ddagger$Dept. of Chemistry, University of Cambridge, Lensfield
Road, CB2 1EW, Cambridge, UK}

\date{\today}
\begin{abstract}
We computed the phase-separation behavior and effective interactions
of colloid-polymer mixtures in the ``protein limit'', where the
polymer radius of gyration is much larger than the colloid radius.
For ideal polymers, the critical colloidal packing fraction tends to
zero, whereas for interacting polymers in a good solvent the behavior
is governed by a universal binodal, implying a constant critical
colloid packing fraction. In both systems the depletion interaction is
not well described by effective pair potentials but requires the
incorporation of many-body contributions.

\end{abstract}
\maketitle Adding polymers to suspensions of micro- and nano-particles
induces depletion interactions that profoundly affect their physical
properties. This phenomenon has important scientific and
(bio-)technological applications.  Polymers such as polyethylene
glycol are routinely added to protein solutions to enable protein
crystallization~\cite{Kulk99,Tana02}, a poorly understood process and
of great importance in structural biology~\cite{Durb96}.  In cell
biology depletion interactions are key in the process of
macro-molecular crowding~\cite{Zimm93}.  Food and paint production are
among the industrial sectors where depletion phenomena play a role.

In this Letter we focus on mixtures of hard sphere (HS) colloids with
a radius $R_c$ and non-adsorbing polymers with a radius of gyration
$R_g$, in the regime where $q=R_g/R_c > 1$. This is often called the
nano-particle or ``protein limit'', because in practice small
particles such as proteins or micelles are needed to achieve large
size-ratios $q$.  Whereas the opposite ``colloid limit'' ($q \lesssim
1$) has been well studied, the physics in the protein limit is less
established.  This imbalance is partially due to the lack of well
characterized experimental model systems for the protein limit and
partially to a poor theoretical understanding.  The colloid limit can
be well described within the framework of effective depletion pair
potentials~\cite{Asak54,Poon02}, in contrast to the protein limit,
where the interactions cannot be reduced to a pairwise
form~\cite{Odij97,Sear01}. Nevertheless, for biological and industrial
applications, this regime is at least as important as the colloid
limit.

One of the first theoretical treatments of colloid-polymer mixtures in
the protein limit was by de Gennes~\cite{deGe79a}, who showed that the
insertion free-energy $F_{c}^{(1)}$ of a single hard, non-adsorbing
sphere into an athermal polymer solution scales as
\begin{equation}\label{eq3}
\beta F_{c}^{(1)} \sim (R_c/\xi)^{3-1/\nu}
\end{equation}  when $R_c < \xi$, with the
polymer correlation length $\xi(\phi_p) \sim R_g
\phi_p^{-\nu/(3\nu-1)} \approx R_g
\phi_p^{-0.77}$~\cite{deGe79}. Here, $\beta=1/k_B T$ is the reciprocal
temperature, $\nu \approx 0.59$ is the Flory exponent and $\phi_p=
\rho_p \frac43 \pi R_g^3$ is the polymer volume fraction for a polymer
number density $\rho_p$, so that $\phi_p \approx 1$ at the crossover
from a dilute to a semi-dilute solution~\cite{deGe79}.  The prefactors
can be calculated by the renormalization group (RG)
theory~\cite{Eise00}, yielding: $\beta F_{c}^{(1)} \approx 4.39 \phi_p
q^{-1.3}$ which has been verified by computer simulations for small
$q$~\cite{Loui02a}.  Based on this description of $ F_{c}^{(1)}$, de
Gennes~\cite{deGe79} and Odijk~\cite{Odij97} predicted extensive
miscibility for colloid-polymer mixtures in the large $q$ limit if
$R_c < \xi$.  However, it is well known that protein-polymer mixtures
do phase-separate~\cite{Doub00,Tuin00}.  Recently, Odijk {\em et
al.}~\cite{Wang01} suggested that a poor solvent could facilitate
phase-separation. Sear~\cite{Sear97} altered the form of $F_{c}^{(1)}$
to include effects when $R_c \gg \xi$, and also predicted
phase-separation with a truncated virial theory. The same author
recently proposed an alternative theory~\cite{Sear02} where the
colloids induce depletion attractions between the polymers, leading to
a poorer effective solvent and eventually phase-separation.
Mean-field cell model calculations also predict
demixing~\cite{Scha97}.  Another promising approach uses integral
equation techniques~\cite{Fuch00} to predict spinodal curves and
critical points.  However, all these theories suffer from several
uncontrolled approximations leading to different predictions for the
causes and properties of the phase-separation.  To clarify this
situation, we performed computer simulations with as few simplifying
assumptions as possible, on which we report in this Letter.

We have recently used a coarse-graining technique~\cite{Loui00} to
study the colloid limit, and found quantitative agreement with
experimental fluid-fluid binodals~\cite{Bolh02}, and significant
qualitative differences between interacting (IP) and non-interacting
(NIP) polymers.  Here, we study the same athermal model of HS colloids
and non-adsorbing polymers in the protein limit, and calculate, for
the first time, the full fluid-fluid binodals by direct simulation.
The results for the IP and NIP show even larger qualitative
differences, and many-body depletion interactions must be invoked to
understand the phase-behavior.

The simulation model consists of polymers on a simple cubic lattice
mixed with HS colloids. The interacting polymers in a good solvent are
modeled as self avoiding walks (SAW) of length $L$, which have a
radius of gyration $R_g \sim L^\nu$.  The non-interacting polymers are
modeled as random-walks, for which $R_g \sim L^{0.5}$.  In both models
there is an excluded-volume interaction between the colloidal HS and
the polymer segments.
\begin{figure}
\includegraphics[angle=0,width=8cm]{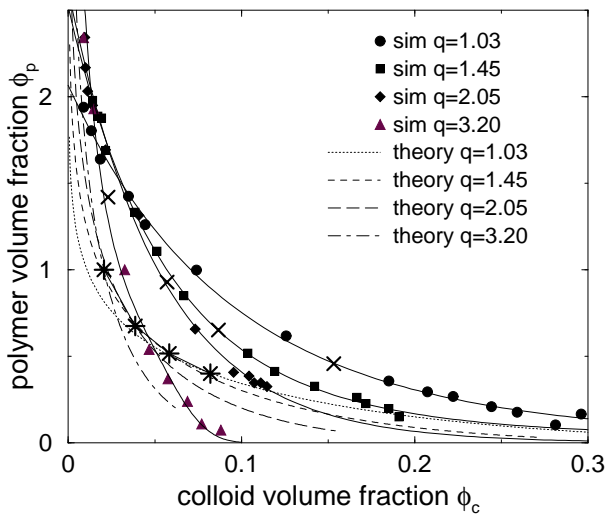}
\caption{Fluid-fluid binodals for a mixture of non-interacting
polymers and HS colloids with different size-ratios $q$.  Crosses
indicate the estimated critical point, obtained by extrapolating the
calculated phase boundaries.  The full lines are a guide to the
eye. Dashed lines denote the simple theory described in the text, with
stars marking the critical points.}
\label{fig:pd_ideal}
\end{figure}
The simulations were performed on a $D^3$ lattice with periodic
boundary conditions, where $D=48$ and $D=100$ for the NIP and IP
system, respectively.  Throughout this Letter, we use the lattice
spacing as the unit of length.  For the NIP the colloidal HS diameter
was $\sigma_{c}=5.5$ and the polymer length was $L=50, 100, 200 $ and
$500$, corresponding to $q=1.03, 1.45, 2.05$ and $3.2$, respectively.
For the IP $L=2000$, and $\sigma_c=10, 14$ and $20$, yielding $q=3.86,
5.58$ and $7.78$, respectively.  Colloidal positions have continuous
values, but when we calculated the interaction between colloid and
polymer the colloids were shifted such that they occupied a constant
number of lattice sites to prevent spurious attractive positions for
single colloids (other lattice effects, although unavoidable, are
expected to be small.)  Thermodynamic state points were calculated in
the grand-canonical ensemble, i.e. at fixed volume $V$, colloid
chemical potential $\mu_{c}$ and polymer chemical potential $\mu_{p}$
using Monte Carlo (MC) techniques.  The NIP were sampled using an
(exact) lattice propagation method~\cite{Fren90,Meij94}, while the IP
configurations were generated using translation, pivot moves and
configurational bias MC~\cite{frenkelbook} in an expanded ensemble to
facilitate insertion of long chains~\cite{Yan00}.  Typical simulations
lengths are $10^9$ Monte Carlo moves per state point.  In order to
determine the liquid-liquid binodals we first estimated the
coexistence line by scanning a series of $\mu_c$ for several values of
$\mu_{p}$ and locate the $\mu_c$ for which a sudden density change
occurred. Subsequently 8-10 $(\mu_p$, $\mu_c)$ coexistence state
points were simulated simultaneously using parallel
tempering~\cite{Yan00}.  When the estimated coexistence points are
sufficiently close to the true binodal and to each other, and near the
critical point, this scheme results in proper ergodic sampling of both
phases.  If necessary, the chemical potentials were adjusted towards
coexistence.  We used the multiple histogram
reweighting~\cite{Ferr87,Yan00} technique to determine the precise
location of the ($\mu_{c},\mu_{p}$) coexistence line, and the phase
boundaries in the ($\phi_c,\phi_p$) plane, where $\phi_c = \rho_c
\frac43 \pi R_c^3$ is the colloid volume fraction, with $\rho_c$ the
colloid number density.

\begin{figure}
\includegraphics[angle=0,width=8.7cm]{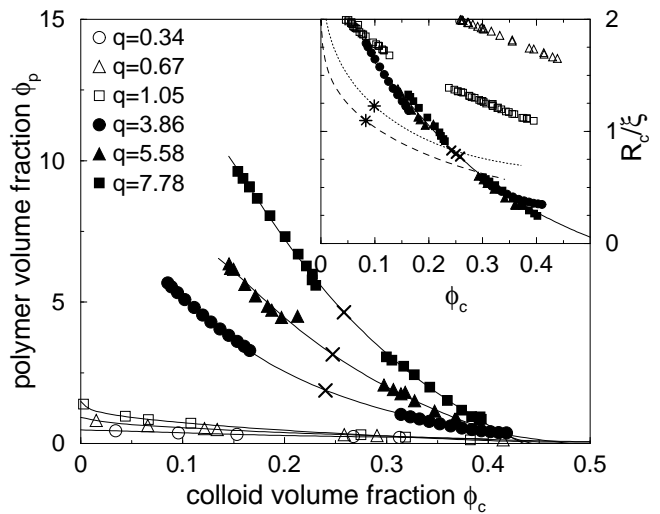}
\caption{Fluid-fluid binodals for a mixture of interacting polymers
and HS colloids at different size-ratios $q$. Filled symbols are
direct simulation data. The open symbols are the colloid limit ($q
\lesssim 1$) results from Ref.~\protect\cite{Bolh02}.  Solid lines are
a guide to the eye.  Inset: The same binodals plotted in a reduced
polymer density representation.  The dotted curve corresponds to a
simple theory for the universal binodal when polymers are in a good
solvent while the dashed line is for polymers in a poorer
solvent. Crosses and stars as in Fig.~\ref{fig:pd_ideal}.}
\label{fig:pd_int}
\end{figure}

Figs.~\ref{fig:pd_ideal} and \ref{fig:pd_int} contrast the calculated
phase diagrams for NIP and IP for several size-ratios $q$.  Firstly,
we note that both models show extensive immiscibility, in agreement
with experiment~\cite{Doub00}.  Secondly, the two systems exhibit
striking differences: for the NIP, the critical colloid volume
fraction $\phi_c^{crit}$ tends to zero with increasing size ratio $q$,
while the IP exhibit a nearly constant value of $\phi_c^{crit}$.  For
both systems the critical polymer concentration $\phi_p^{crit}$
increases with increasing $q$. The phase-separation occurs well into
the semi-dilute regime for the IP, again in qualitative agreement with
experiment~\cite{Tuin00}.  Properties of semi-dilute polymer solutions
are independent of polymer length, being instead determined by the
correlation length $\xi$, which is a function of the monomer density
$c=L\rho_p$.  The phase behavior of the polymer-HS mixture should
therefore only be a function of the ratio $R_c/\xi$~\cite{Sear97}.
Indeed, when the phase lines in Fig.~\ref{fig:pd_int} are rescaled
with an accurate prescription for $\xi(\rho_p)$~\cite{Loui02a}, the
binodals nearly collapse onto a ``universal binodal'', as shown in the
inset of Fig.~\ref{fig:pd_int}.  This explains why the critical
colloid packing fraction is nearly constant in the simulations.
Similarly, $\phi_p^{crit}$ scales as $\phi_p^{crit} \sim
q^{(3\nu-1)/\nu} \approx q^{1.3}$.  For comparison, we have also
included results for $q \lesssim 1$ from Ref.~\cite{Bolh02} in
Fig.~\ref{fig:pd_int}.  These results do not exhibit the same scaling
behavior, since they are not in the semi-dilute regime.

The differences between NIP and IP phase behavior can be rationalized
with some simple theories.  Consider a Helmholtz free-energy $F$ of
the form $\beta F/V = f = f_c^{HS} + f_p + f_{cp}$.  Here, the HS free
energy $f_c^{HS}$ is given by the accurate Carnahan-Starling
expression~\cite{Hans86}, and the polymer free energy $f_p$ for either
IP or NIP solutions is well understood~\cite{deGe79}.  The
contribution due to the HS-polymer interactions $f_{cp}$ is
non-trivial. A first approximation truncates after the second
cross-virial coefficient, yielding $f_{cp} \sim \rho_c F_c^{(1)}$.
For NIP the insertion free-energy $F_c^{(1)}$ is exactly
known~\cite{Eise00}, so that $f_{cp}$ takes the form $f_{cp}^{id} =
\rho_p \phi_c \left(1 + \frac{6 q }{\sqrt{\pi}} + 3 q^2 \right) \equiv
\rho_p \phi_c \hat{b}_{cp}$, which defines the reduced cross-virial
coefficient $\hat{b}_{cp}$. Since $f_{cp}^{id}$ grows with increasing
$q$, immiscibility sets in at lower colloid packing fraction $\phi_c$.
The theory can be improved by realizing that the polymers only exist
in the free volume left by the colloids~\cite{Lekk92}.  Simply taking
this free volume to be $1-\phi_c$ is an adequate first approximation
for the protein limit.  The trends for the binodal lines calculated
from this simple theory, shown in Fig.~\ref{fig:pd_ideal}, agree
qualitatively with the simulations. For example, the critical point
shifts to smaller $\phi_c$ and larger $\phi_p$ for increasing $q$, and
the binodal lines cross at a low $\phi_c$.  For computational reasons
the simulations only go up to $q =3.2$ and we expect better
quantitative agreement for larger $q$ since $\phi_c^{crit}$ decreases
so that the second-virial theory should become more accurate.  In the
$q \rightarrow \infty$ limit, this theory yields $\phi_c^{crit}
\rightarrow 1/\hat{b}_{cp} \sim 1/(3q^2)$, and $\phi_p^{crit} =
q^3/\hat{b}_{cp} \sim q/3$.  Note that in the same limit, the
penetrable sphere or Asakura-Oosawa model~\cite{Lekk92} scales
somewhat differently: $\phi_c^{crit} \rightarrow 1/q^3$ and
$\phi_p^{crit} \rightarrow 1$.  Sear~\cite{Sear01} already pointed out
the $\phi_c^{crit} \rightarrow 0$ behavior using a slightly different
prescription for the free volume than we employ here.  Here we claim
that the limiting results are a general feature of free-volume
theories.

In the IP case, $f_{cp}$ is more difficult to estimate, even for a
second cross-virial theory.  The $R_c \ll \xi$ limit is given by
Eq.~(\ref{eq3}) with the prefactors from RG theory.  For $R_c \geq
\xi$ we have previously shown that $F^{(1)}_{c}$ is given by $
F^{(1)}_{c} = \frac43 \pi R_c^3 \Pi + 6 \pi R_c^2
\gamma_s$~\cite{Loui02a}, where the polymer osmotic pressure $\Pi \sim
\xi^{-3}$ is well known~\cite{deGe79}.  However, since Eq.~(\ref{eq3})
is essentially a surface (depletion layer) contribution, we use a
simple approximate second cross-virial term $ f_{cp} = \rho_c \left(
\beta \Pi(\rho) \frac43 \pi R_c^3 + 4.39\phi_p q^{-1.3} \right)$,
which reduces to the correct form in both the $R_c \ll \xi$ and the
$R_c \gg \xi $ limit~\footnote{Obviously this approach, which
resembles the more schematic theory of Sear~\cite{Sear97}, could be
improved.}.  As with our treatment of NIP, we take the effect of the
colloid excluded volume into account by computing the polymer
densities in the free volume fraction $1- \phi_c$ (see
Ref.~\cite{Aart02} for a complimentary approach).  The theoretical
binodals were calculated using accurate expressions for $\xi$ and
$\Pi$~\cite{Loui02a} and are compared with the simulation results in
Fig.~\ref{fig:pd_int}, in the $R_c/\xi$ versus $\phi_c$ plane.  The
qualitative agreement suggests that we can also use this theory to
estimate the effect of a poorer solvent on the binodals.  Following
Ref.~\cite{Wang01}, we alter the scaling of $\xi$ to $\xi \sim
\phi_p^{-\delta/3}$ so that $\Pi \sim \xi^{-3} \sim \rho_p
\phi_p^{\delta-1}$.  Interestingly, Fig.~\ref{fig:pd_int} shows that
using $\delta \sim 1.5$ instead of the appropriate exponent for
polymers in a good solvent ($\delta \sim 2.3$), does not result in
important differences in the binodals.  Of course, the differences
will appear larger in the ($\phi_c,\phi_p$) plane due to the different
scaling of $\xi$.  One must keep in mind, however, that these
predictions follow from a simple scaling theory and qualitatively
different behavior may emerge when one approaches the $\theta$-point
(where $\delta = 1$).

\begin{figure}[b]
\includegraphics[angle=0,width=8.3cm]{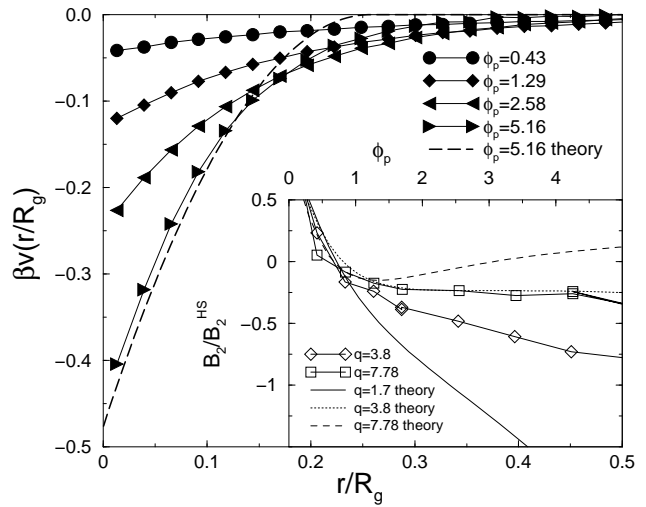}
\caption{ Effective colloid-colloid pair potentials induced by
interacting polymers for q=7.78. Theoretical lines from
Ref.~\protect\cite{Loui02b}.  Inset: Reduced second osmotic virial
coefficient $B_2^* = B_2/(\frac23 \pi R_c^3)$ as a function polymer
densities for several size ratio's.}
\label{fig:v2}
\end{figure}

To illustrate the many-body nature of the depletion interaction we
estimated the phase behavior by approximating the system by colloids
interacting via pairwise effective potentials. We computed the
effective pair interaction $v(r)$ between two colloids in a bath of
IP's, by $\beta v(r)=-\ln g(r)$ for $\rho_c \rightarrow 0$.  The
colloid radial distribution function $g(r)$ was estimated by measuring
the insertion probability of a HS at a distance $r$ from a second
fixed HS in a SAW polymer solution using the above MC techniques.
Results for a single size-ratio $q=7.78$ as a function of $\phi_p$ are
shown in Fig.~\ref{fig:v2}.  Several features are similar to the
colloid limit~\cite{Loui02b}: the range shortens and the well-depth
increases with increasing $\phi_p$.  Interestingly, our simple
depletion potential~\cite{Loui02b}, derived for the colloidal limit,
also works semi-quantitatively in this regime.  A good measure for the
attractive strength of an effective pair potentials is given by the
second osmotic virial coefficient~\cite{Hans86}, shown in
Fig.~\ref{fig:v2}. The saturation of $B_2^*$ for larger $q$ is an
interesting qualitative feature: apparently the shortening of the
range compensates the deepening of the attraction, so that the total
cohesion does not increase with increasing $\phi_p$, something also
found in RG~\cite{Eise00} and integral equation
calculations~\cite{Fuch00}.  For pairwise interacting systems,
phase-separation typically sets in when $B_2^* \leq
-1.5$~\cite{Vlie00}. Here, the saturation of $B_2^*$ suggests that for
large $q$ the pair interactions do not provide enough cohesion to
explain the phase-separation.  We arrive at the same conclusions with
simple mean-field theories~\cite{Hans86}, which should be relatively
trustworthy given the long range of the pair potentials.  Obviously,
for $q\gg 1$ a pair-level description is not sufficient, and many-body
interactions must be invoked.

For the NIP, one might also expect many-body interactions to be
important for large $q$.  A good approximation to the pair-potentials
exists~\cite{Meij94,Loui02b}, from which the second virial
coefficients at the calculated critical points follow:
$B_2^*(q=1.03)\approx -13.1$; $B_2^*(q=1.45)\approx -16.4$;
$B_2^*(q=2.05)\approx -22.7$.  Even though the actual critical
$\phi_c$'s are very low, so that a second-virial description might be
thought to be sufficient, the analysis above shows that for NIP the
pair-interactions provide too much cohesion, {\em opposite} to the IP
case. Clearly, many-body interactions must also be invoked to describe
the phase-behavior correctly, as suggested by other
authors~\cite{Gast83, Meij94,Odij97, Fuch00,Sear01}.

In conclusion, we have shown by computer simulations that a mixture of
polymers and non-adsorbing HS colloids shows extensive immiscibility
in the protein limit, where the polymer-colloid size-ratio $q \gg 1$.
For IP the phase-behavior is dictated by a universal binodal in the
semi-dilute regime.  For NIP, the colloid packing fraction tends to
zero for increasing polymer length.  In contrast to the better studied
colloid limit, pair interactions are not sufficient to rationalize the
phase behavior.  We hope that future experiments on HS colloids with
non-adsorbing polymer will test these predictions.  Future work might
include extensions to non-spherical particles, poor solvents, and
adsorbing systems.

\begin{acknowledgments}
A.A.L. acknowledges the Isaac Newton Trust for financial support.
E.J.M. acknowledges the Royal Netherlands Academy of Art and Sciences
for financial support.  We acknowledge support from the Stichting
Nationale Computerfaciliteiten (NCF) and the Nederlandse Organisatie
voor Wetenschappelijk Onderzoek (NWO) for the use of supercomputer
facilities. We thank M. Fuchs, R. Sear and R. Tuinier for helpful
discussions.
\end{acknowledgments}


\begin{thebibliography}{99}

\bibitem{Kulk99} A. M. Kulkarni {\em et al.}, Phys. Rev. Lett. {\bf
83}, 4554 (1999); J. Chem. Phys. {\bf 113}, 9863 (2000).
\bibitem{Tana02} S. Tanaka and M. Ataka, J. Chem. Phys. {\bf 117},
3504 (2002).
\bibitem{Durb96} S.D. Durbin and G. Feher, Annu. Rev. Phys. Chem. {\bf
47}, 171 (1996).
\bibitem{Zimm93} S.B. Zimmerman and A.P. Minton,
Annu. Rev. Bioph. Biom. {\bf 22}, 27 (1993).
\bibitem{Asak54} S. Asakura and F. Oosawa, J. Chem. Phys. {\bf 22},
1255 (1954).
\bibitem{Poon02} See W.~C.~K. Poon, J. Phys.: Condens. Matter {\bf
14}, R859 (2002) for a recent review.
\bibitem{Sear01}R.P. Sear, Phys. Rev. Lett {\bf 86} 4696 (2001).
\bibitem{Odij97} T. Odijk J. Chem. Phys. {\bf 106} 3402 (1997).
\bibitem{deGe79a} P.G. de Gennes, C. R. Acad. Sci., Paris B {\bf 288},
359 (1979).
\bibitem{deGe79} P.G de Gennes, {\em Scaling Concepts in Polymer
Physics}, (Cornell University Press, Ithaca).
\bibitem{Eise00} E. Eisenriegler {\em et al.}, Phys. Rev. E {\bf 54},
1134 (1996); E. Eisenriegler, J. Chem. Phys. {\bf 113}, 5091 (2000).
\bibitem{Loui02a} A.A. Louis, P.G. Bolhuis, J.P. Hansen, and
E.J. Meijer, J. Chem. Phys. {\bf 116}, 10547 (2002).
\bibitem{Tuin00}R. Tuinier {\em et al.}, Langmuir {\bf 16}, 1497
(2000).
\bibitem{Doub00}For a review see e.g. J.-L. Doublier {\em et al.},
Curr. Opin. Colloid In. {\bf 5 }, 202 (2000).
\bibitem{Wang01} S. Wang {\em et al.}, Biomacromolecules {\bf 2}, 1080
(2001).
\bibitem{Sear97} R.P. Sear, Phys. Rev. E {\bf 56}, 4463 (1997).
\bibitem{Sear02} R.P. Sear, preprint, cond-mat/0206320.
\bibitem{Scha97} H.M Schaink and J.M Smit, J. Chem. Phys {\bf 107},
1004 (1997).
\bibitem{Fuch00} M. Fuchs and K.S. Schweizer, Europhys Lett. {\bf 51},
21 (2000); J. Phys.: Condens. Matt. {\bf 14}, R239 (2002).
\bibitem{Loui00} A.A. Louis {\em et al.}, Phys. Rev. Lett. {\bf 85},
2522 (2000); P.G. Bolhuis {\em et al.}, J. Chem. Phys. {\bf 114}, 4296
(2001).
\bibitem{Bolh02} P.G. Bolhuis {\em et al.},Phys. Rev. Lett. {\bf 89}
128302 (2002).
\bibitem{Fren90} D.~Frenkel, J.  Phys.: Condens. Matter {\bf 2}, SA265
(1990).
\bibitem{Meij94} E.J. Meijer and D. Frenkel, Phys. Rev. Lett. {\bf
67}, 1110 (1991); J. Chem. Phys.  {\bf 100}, 6873 (1994).
\bibitem{frenkelbook}D. Frenkel and B. Smit, {\it Understanding
molecular simulations, 2nd ed.} (Academic Press, San Diego, 2002).
\bibitem{Yan00} Q.~Yan and J.~J.~de~Pablo, J. Chem. Phys. {\bf 113},
1276 (2000).
\bibitem{Ferr87}A.~M. Ferrenberg and R.~H.~Swendsen,
Phys. Rev. Lett. {\bf 61}, 2635 (1987).
\bibitem{Hans86} J.P. Hansen and I.R. McDonald, {\em Theory of Simple
Liquids, 2nd Ed.} (Academic Press, London, 1986).
\bibitem{Lekk92} H.N.W. Lekkerkerker {\em et al.},
Europhys. Lett. {\bf 20}, 559 (1992).
\bibitem{Aart02}D. G. A. L. Aarts {\em et al.}, J. Phys.:
Condens. Matter {\bf 14}, (2002).
\bibitem{Loui02b} A.A. Louis {\em et al.}, J. Chem. Phys. {\bf 117},
1893 (2002).
\bibitem{Vlie00} G. A. Vliegenthart and H. N. W. Lekkerkerker,
J. Chem.  Phys. {\bf 112}, 5364 (2000).
\bibitem{Gast83}A.P. Gast, C.K. Hall and W.B. Russel,
J. Colloid Interf. Sci. {\bf 96}, 251 (1983).
\end{thebibliography}
\end{document}